# Error-rate reduction in network-based biocomputation

*Pradheebha Surendiran[1,2], Marko Ušaj[3], Till Korten[4], Alf Månsson[1,3], Heiner Linke[1,2\*]*

1 NanoLund, Box 118, Lund University, 22100 Lund, Sweden

2 Solid State Physics, Box 118, Lund University, 22100 Lund, Sweden

3 Department of Chemistry and Biomedical Sciences, Linnaeus University, Kalmar, Sweden

4 B CUBE - Center for Molecular Bioengineering, Technische Universität Dresden, D-01307 Dresden, Germany

**Abstract**

Network-based biocomputation (NBC) is an alternative parallel computing paradigm that encodes combinatorial problems into a nanofabricated device's graphical network of channels, enabling cytoskeletal filaments propelled by molecular motors to explore the problems' solution space. NBC promises to require significantly less energy than traditional computers due to the high energy efficiency of molecular motors. However, error rates associated with the pass junction crossing, the primary path-regulating geometry, pose a bottleneck for scaling up this technology. Here, we optimize the geometry of the pass junction for low error rates for the actin-myosin system. To do so, we evaluate various pass junction designs that differ in features, such as the nanochannel width, junction crossing area, and angles of a funnel-shaped output part of the junction. Error rates were measured experimentally by using gliding motility assay and as well as by simulation methods. The final optimized design displayed a decreased error rate of under 1% compared to the previous 2-4%. We anticipate this improvement will enable scaling up NBC networks from tens to hundreds of pass junctions. However, the results of 2D junction optimizations also suggest that further drastic reduction of error rates in two-dimensional pass junctions is unlikely, necessitating three-dimensional junctions, such as bridges or tunnels, for complete error rate mitigation. Furthermore, the simulation results demonstrate that including a layer of myosin motor on the channel provides a better fit between simulation and experimental results.

**Keywords:** biocomputation, actin-myosin, nanofabrication, junctions, motility, simulation

**Introduction**

To solve demanding combinatorial problems such as non-deterministic polynomial time (NP)-complete problems, alternate parallel computing technologies have been developed where many possible solutions are processed simultaneously to reduce processing time or energy consumption[1,2]. Network-based biocomputation (NBC) is an emerging alternative computing approach that encodes a given mathematical problem into a graphical nanofabricated network, which is subsequently explored in parallel by molecular motor-propelled filaments. Acting as molecular computing agents, the filaments exhibit the ability to navigate through the intricate network structure, utilizing chemical energy in the form of adenosine triphosphate (ATP) to generate mechanical movement. NBC shows the potential to significantly reduce energy consumption[2], compared to traditional electronic computers[1,3], enabled by (i) the energy efficiency of molecular motors[4] and (ii) the inherent thermodynamic advantage of the parallel approach[5]. NBC has been used to solve the subset sum problem (SSP) with eight possible solutions[1], an Exact Cover of 32 and 1024 possible solutions[3], and the 3-Satisfiability problem with up to three variables and five clauses[6].

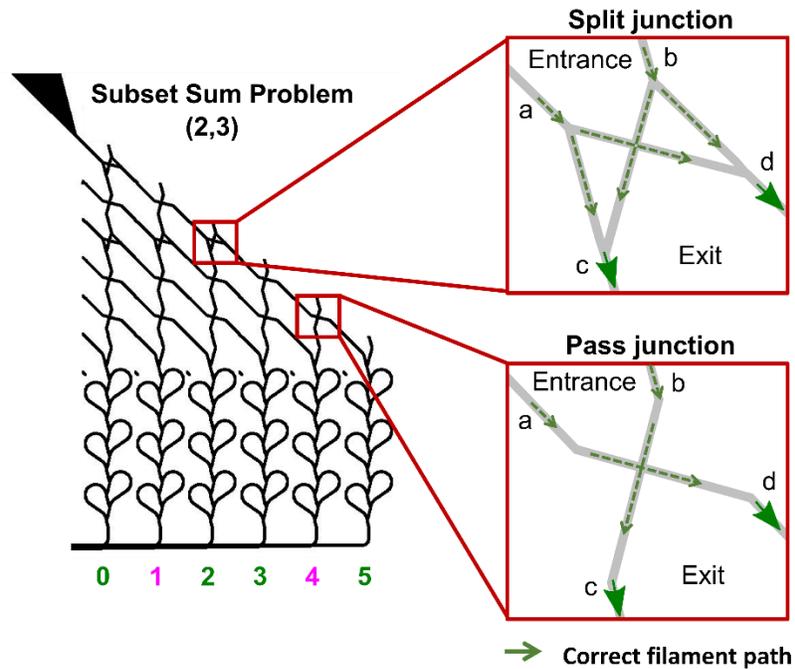

**Figure 1.** Illustration of an NBC device solving SSP (2, 3) with the main geometries, split junction, and pass junction. The green arrows in the insets represent the correct (intended) filament paths within these junctions. Entrance channels are labeled a and b. Exit channels are labeled c and d respectively. In the SSP network (main panel), the exit numbers correspond to the target sums; the green- and magenta-colored numbers show the correct and incorrect solution exits respectively.

We illustrate how NBC works using an SSP network as shown in Figure 1. Two types of geometries regulate the movement of agents within the network. The agents enter the network from the top-left corner and encounter split junctions, where they ideally have an equal probability of either continuing straight ahead or turning. Pass junctions, on the other hand, are intended to only allow the agents to move straight ahead. The correct, expected filament paths through the split and pass junctions are indicated as green arrows in Figure 1. If an agent moves diagonally down at a split junction, it adds an integer to its position, where the value of this integer is determined by the number of rows of pass junctions until the next split junction.

Errors at pass junctions have been identified as one of the key limiting factors for NBC scalability.[7,8] If an agent at one of the pass junctions takes a path other than the allowed direction indicated, this agent will end up at an incorrect solution exit and produce an erroneous count at that exit (Figure 1). The junction error rate $E$ is calculated as the total number of incorrect paths at a junction divided by the total number of passing filaments, $N$. The number of erroneous counts scales with $(1 - E)^x$, where $x$ is the number of pass junctions crossed.

This study integrates experimental investigations and simulation to optimize the pass junction geometry for a small pass junction error rate $E$, specifically for the actin-myosin system. By closely examining filament behavior that led to errors in previous experiments and simulations, a series of trial junction designs was developed. We evaluate the defined geometries through experiments and simulation to measure and analyze the resulting error rates. The error rates are quantified using a small-scale instance of the SSP (2,3) network depicted in Figure 1. An optimized junction geometry was successfully identified in the experiments, demonstrating an error rate of less than 1%. Additionally, Monte Carlo simulations were employed to simulate filament movement within the designed junctions. The experimental and simulation results suggest that the channel widths used in simulations to account for a myosin motor layer improves the agreement with experiment.

**Results and discussion**

*Approach for optimization of junction geometry*

In previous studies using the actin-myosin system within network-based biocomputation (NBC), two types of pass junctions have been utilized: a simple straight pass junction[1] and a pass junction with a funnel[3]. These junctions were found to yield $E$ = 2.1% and 3.8 %, respectively. In this study, we aim to identify the key parameters that contribute to $E$ depending on the junction type employed.

Specifically, we tune three geometric features to find an optimal junction geometry. (1) The width of the channels relative to the filament persistence length has been previously identified as important for the behavior at junctions[9,10]. We explore the range from channel width $C$ = 50 nm (the smallest where we observed reproducible filament transport) up to $C$ = 200 nm as the filaments were shown to take U-turns for $C$ > 200 nm. (2) Previous observations[11] show that filaments travel along the edge in the incoming channels with the risk that their tips may hit the wall of the perpendicular channel rather than moving into the appropriate outgoing channels. To mitigate this risk, we introduce offsets ranging from 0 to 50% of the total channel width. (3) The inclusion of funnels in the geometry has been observed to enhance filament collection and facilitate movement along the correct paths[1].

Based on these observations, we developed a series of designs that incorporated parameter values as depicted in Figure 2 and Table 1. These designs focused on the manipulation of key parameters, namely channel width (C), offset distance (O), funnel width (F), and funnel angles (α and β), with each parameter varied individually while keeping the others constant. An overview of the parameters for all the designs can be found in Table 1.

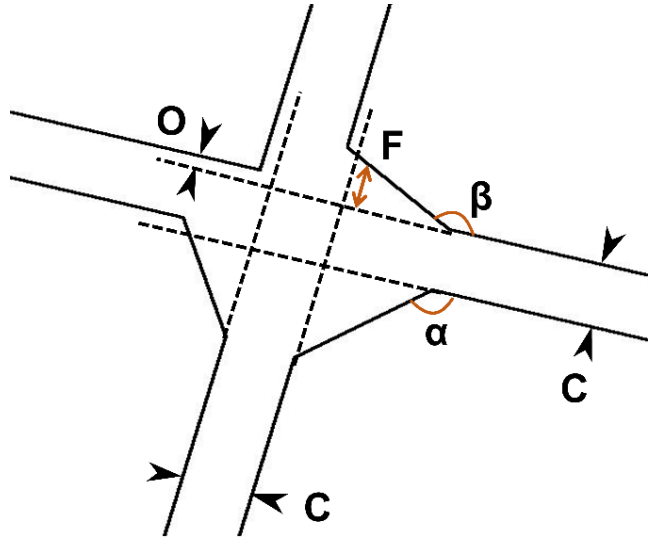

**Figure 2.** Schematic diagram of the pass junction showing the parameters of the geometry.

**Table 1.** Summary of key designs of pass junctions with relevant parameters

| Parameter annotations | Design Name | C (nm) | O (nm) | F (nm) | α | β |
|---|---|---|---|---|---|---|
| C (Channel width) | $C_{50}$ | **50** | 0 | 30 | 45 | 30 |
| | $C_{100}$ | **100** | 0 | 30 | 45 | 30 |
| | $C_{150}$ | **150** | 0 | 30 | 45 | 30 |
| | $C_{200}$ | **200** | 0 | 30 | 45 | 30 |
| O (Offset distance) | $O_0$ | 100 | **0** | 25 | 45 | 30 |
| | $O_{15}$ | 100 | **15** | 25 | 45 | 30 |
| | $O_{25}$ | 100 | **25** | 25 | 45 | 30 |
| | $O_{45}$ | 100 | **45** | 25 | 45 | 30 |
| F (Funnel width) | $F_0$ | 100 | 22 | **0** | - | - |
| | $F_{25}$ | 100 | 22 | **25** | 45 | 30 |
| | $F_{50}$ | 100 | 22 | **50** | 45 | 30 |
| | $F_{110}$ | 100 | 22 | **110** | 45 | 30 |
| α, β (Funnel inner and outer angles) | $\alpha_0\beta_0$ | 100 | 0 | 0 | - | - |
| | $\alpha_{30}\beta_{30}$ | 100 | 0 | 30 | **30** | **30** |
| | $\alpha_{45}\beta_{45}$ | 100 | 0 | 55 | **45** | **45** |
| | $\alpha_{45}\beta_{30}$ | 100 | 0 | 30 | **45** | **30** |

*Experimental evaluation*

Our approach is to evaluate the error rates of the junction designs within an actual SSP network as illustrated in Figure 1. The fabrication of the devices was carried out using electron beam lithography (see Methods). For the experimental setup, actin filaments propelled by myosin II motor fragments (heavy meromyosin, HMM) were employed as biological agents[12]. The operation of the devices was conducted using an in-vitro motility assay[13] where fluorescently labeled actin filaments are guided by surface-adsorbed, non-labeled myosin-II motor fragments to navigate through the network structure.

To analyze the error rates, 1500 frames of 0.2 s fluorescence time-lapse movies of the process were recorded (details in the Methods). The movement of filaments in all the pass junctions was evaluated to find the number of correct and incorrect filament paths. Filaments adhering to the intended routes within the pass junctions are termed correct filaments, while those deviating from the desired paths are categorized as incorrect filaments. The junction error rate $E$ was calculated as the ratio of filaments taking incorrect turns in the pass junctions to the total number of filaments that crossed the junction. Measured error rates for all experimental designs are shown in Figure 3. These data show that designs featuring lower channel widths, lower offset distances, and higher funnel widths tend to show lower error rates. The lowest error rate (0.9%) is observed for the $F_{110}$ design.

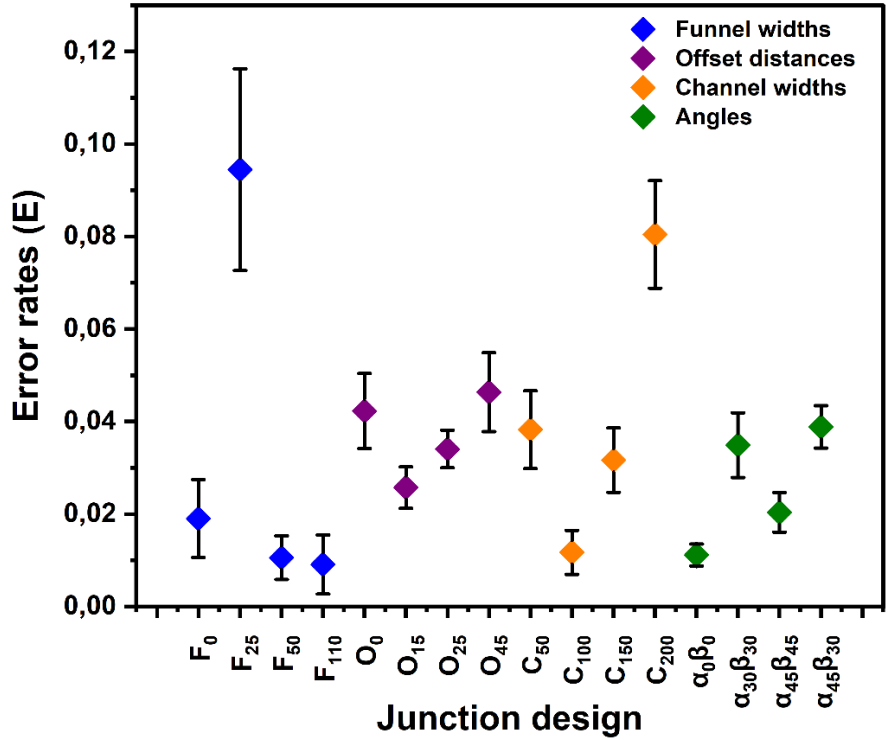

**Figure 3.** The experimentally observed error rates (E) for the 16 designs (Table 1) tested. Error bars represent the counting error ($\frac{\sqrt{E*(1-E)}}{N}$).

### Simulated filament behavior in different surface geometries.

We employed a Monte-Carlo simulation approach, developed from previous work[1,11,14] and implemented using MATLAB (Mathworks, Natick, MA) (see Supplementary section 1 for detailed description). Actin filament movement was modeled as a two-dimensional walk[11] within a nanofabricated channel, characterized by a specific persistence length of the filament and confinement within a predefined region of the surface[9]. Briefly, each filament is modeled as a point, representing the position of its tip. The software mimics the trajectories of cytoskeletal filaments as they traverse through guiding channels, which entails tracking the positional changes of their tips at specific time intervals[15]. When a filament encounters a wall, it becomes oriented along the wall's direction as shown in Figure 4(A), and the extent to which it diverges from its original path is determined by its flexural rigidity, proportional to the persistence length ($L_p$). One parameter that could influence the behavior of the filament in the channels would be edge stickiness (ES). This simulation feature is inspired by the observations in experiments where filaments, when they hit the channel wall at sharp angles, get stuck and turn around. ES is simulated as a probability of the filament getting stuck on the wall and turning around as shown in Figure 4(B).

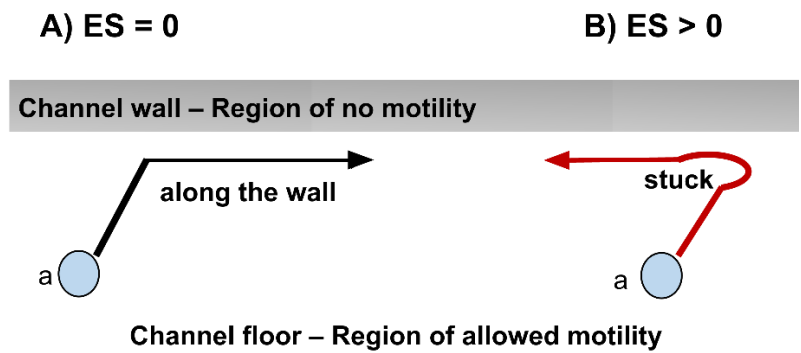

**Figure 4.** (A) Simulation model. The movement of the filament tip (blue filled circle) is simulated in a channel. When the filament tip reaches a channel wall, the filament aligns itself along the wall. Between each time step, a variation is added that is quantitatively described by thermal fluctuations corresponding to a given $L_p$-value (see SI 1.1). (B) When the edge stickiness parameter increases, the filament has a probability > 0 to get stuck on the wall and curl to turn around.

To investigate the behavior of filaments within the designed junctions, simulations were initially performed using a model with $L_p$ = 10 µm and ES = 0. These simulations yielded heatmaps and movies showing the frequently visited pixels of the filaments, depicted in bright colors (Figure 5, represented by blue in this study). To gain insights into the fundamental behavior, we examined heat maps of designs with the lowest and highest parameter values. The following observations were made in simulation for each set of designs in Figure 5:

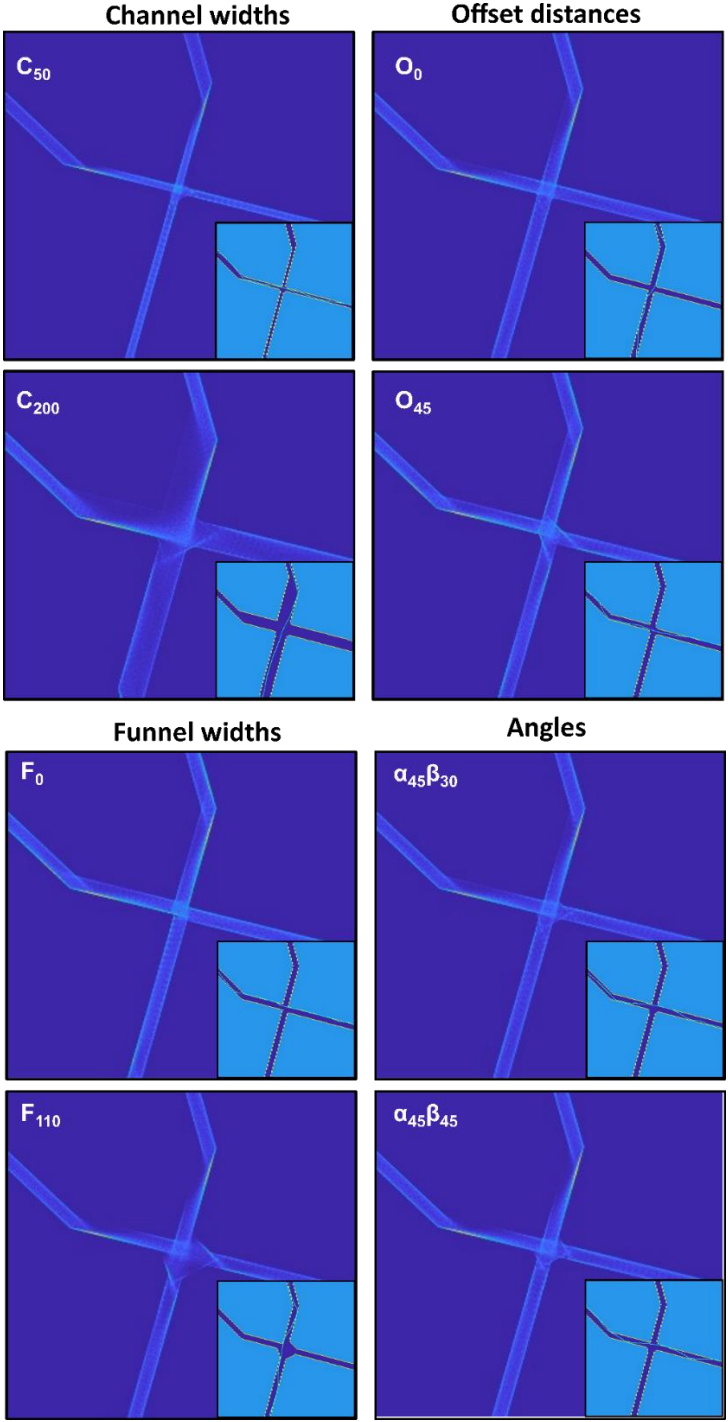

**Figure 5.** Surface and heatmaps of the junction geometries. The heatmaps on the right indicate the frequency at which pixels were visited by simulated filament tips, the brighter a pixel appears, the more frequently it was visited by a filament tip. In the insets, the blue lines indicate an example path taken by the filaments for the respective designs obtained from the simulation data.

**Channel widths (C in Figure 5).** Analysis of the simulation results involved the examination of heatmaps and movies to gain insights into filament behavior under the conditions assumed in the simulations. The narrowest $C_{50}$ design ($C = 50$ nm), facilitates the correct movement of filaments due to the restricted channel space, thereby leading to filaments quickly moving across the channel and continuing the directed path. Conversely, designs featuring wider channels such as the $C_{200}$ design ($C = 200$ nm), give the filaments more space to take sharper turns in the junction with high possibilities of errors (see Supplementary video S1).

**Offsets (O in Figure 5).** Offsets were introduced because we observed that more filaments move along the bottom wall of the left entrance channel and at the right wall of the top entrance channel as depicted by the brighter pixels in the heatmaps shown in Figure 5. This is because of the kink in the channel directly before the junction in the SSP network design. Analysis of the simulation heatmaps indicated that designs featuring a high offset O = 45 nm (designs named $O_{45}$), of nearly half the channel width $C = 100$ nm, exhibit high error rates attributed to a high likelihood of corner collisions. Whereas designs with a lower offset O = 15 nm (designs named $O_{15}$) performed better with lower error rates.

**Funnel widths (F in Figure 5).** The range of funnel widths examined was $0 < F < 110$ nm. Excessively wide funnels provide ample space for filaments to deviate within the center of the junction. The implementation of funnels proved to be a useful strategy for collecting and guiding filaments in conjunction with offset distances. Comparing the heatmaps of two junction designs, featuring the same offset distance and channel widths, we observed that the design without a funnel ($F_0$) resulted in filament collisions with sharp corners, leading to errors. Conversely, designs with F = 110 nm (designs named $F_{110}$) successfully collected and directed filaments toward the straight path (see Supplementary video S2).

**Angles ($\alpha$ and $\beta$ in Figure 5).** The shape of the funnels can significantly impact filament guidance and collection, and this is modulated by varying the angles of the funnel design. The funnel in the junctions acts as a guiding structure where the filaments hit, and due to their angles, the filaments can slide into the outgoing junction and prevent it from taking the wrong direction. Various angles at the junctions determine how filaments interact when they collide within the funnels and corners. Analysis of the heatmaps in the simulations indicated that $\alpha_{45}\beta_{45}$ ($\alpha$ = 45°; $\beta$ = 45°) design depicted in the figure showed reasonably effective guidance compared to others.

*Comparison of experiment and simulation error rates*

In this section, we compare the error rates observed in the experimental data with those obtained through simulations. We simulated the designs used for experiments with $L_p$ = 10 μm[11,16]. Use of ES > 0 was necessary as filaments getting stuck and turning around was observed at times in our experiments. Previous literature suggests that HMM would be expected to form a dense layer of 25 nm thickness or more on the channel walls[17]. Here we include this layer into our simulations and consider three possibilities to understand the impact of the mentioned HMM layer on error rates: (i) the design geometry which was used in the fabrication of the experimental devices, (ii) modified geometry where only one entrance channel width is reduced compared to the design, considering HMM layers and (iii) modified geometry where all the channels were narrowed assuming that filaments would be unable to penetrate the HMM layer.

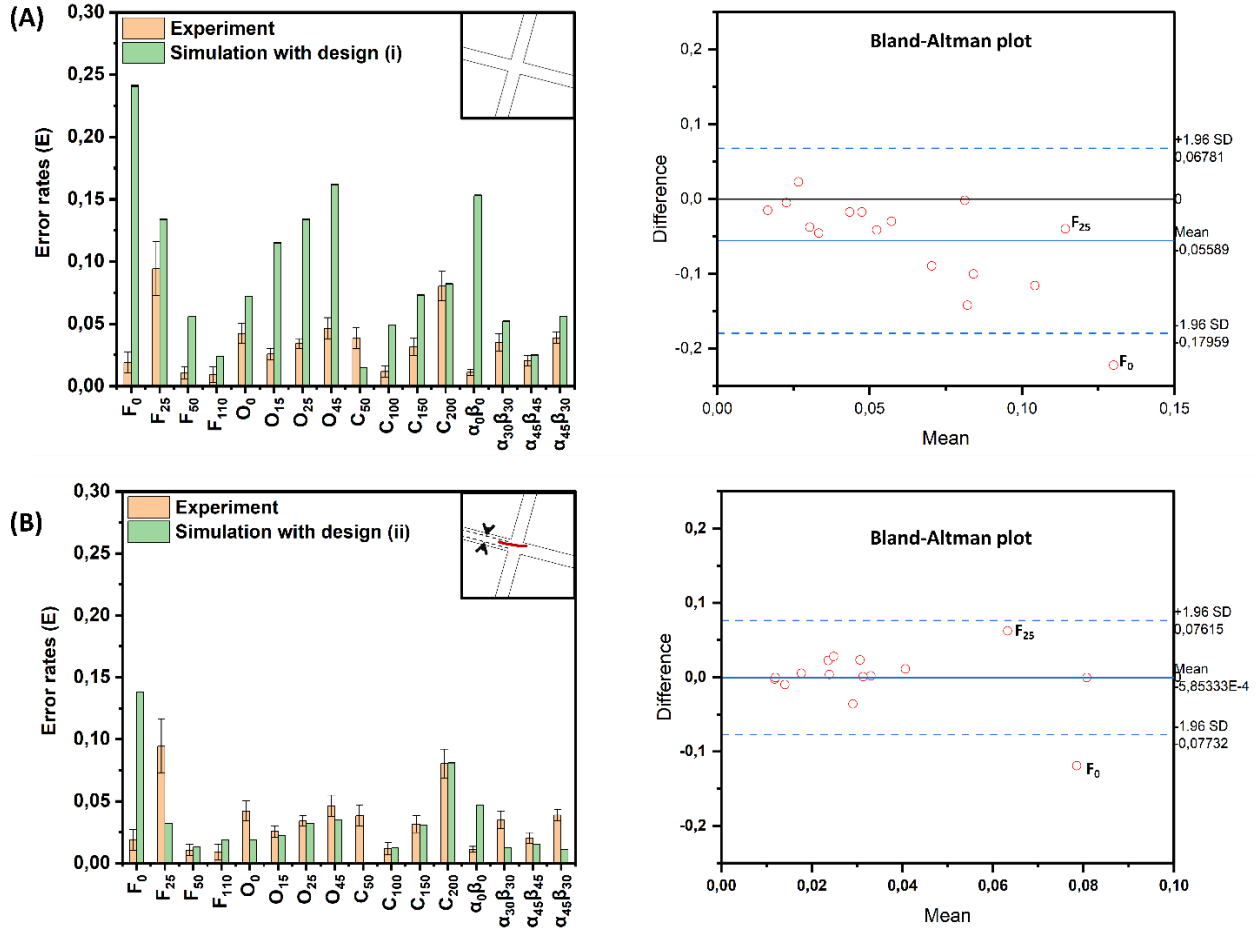

**Figure 6.** Comparison of experimental and simulation error rates in absolute values (left) and using Bland-Altman plots (right). (A) Simulation results based on design geometry (i) used for experiments as exemplified in the inset. (B) Simulation results of design geometry (ii) with reduced width of incoming channels shown by black arrows in inset to account for the heavy meromyosin HMM layer (dashed lines in inset). Filament (red-colored in inset) penetration of the HMM layer in the channel after the junction is depicted (gliding direction from left to right). Error bars in main panels represent the counting error ($\frac{\sqrt{E*(1-E)}}{N}$). Here, HMM thickness was assumed to be 30 nm and ES = 0.12 in the simulations. In Bland–Altman plots, the horizontal axis depicts the overall mean value of $E$ measured by experiment and simulation, and the vertical axis

represents the difference of measured values. Solid color blue line denotes the bias and dashed lines 95% upper and lower limits of agreement.

The design (ii), is based on the hypothesis that the actin filament, gliding along the entrance channel does not penetrate the layer of HMM because it is interacting along its entire length with this layer. However, upon entering the junction the filament tip (diameter approximately 10 nm) is expected to penetrate the HMM layer in all three outgoing channels (see detailed description in Supplementary section 2 and Figure S2). To assess this hypothesis, we evaluated the agreement between experimental and simulation error rates of the two designs (i) and (ii) using Bland- Altman (BA) plot[18], aiming to identify the conditions that closely approximate the experimental results. To refine the model and ensure its alignment with experimental conditions, simulations were conducted by varying the thickness of the heavy meromyosin (HMM) layer within the range of 25-35 nm[17], along with different edge stickiness. To compare simulation and experiment we use a BA plot, to measure the agreement between the two techniques by plotting the mean and difference of the measured values (for a detailed description, see Supplementary section 4)[19]. The BA plot visually represents the mean difference (bias), shows the spread of differences in a scatter plot, and provides limits of agreement to quantify the range within which most differences between the methods lie.

The BA plots (Figure 6) suggest that the simulations using geometry (i), that do not account for any HMM layer, did not agree well with the experimental data: the mean difference was significantly different than zero and the values exhibit a large degree of scatter. (Figure 6(A)). The BA plot of geometry (ii) in contrast shows a mean difference close to zero (Figure 6B), and most data points cluster tightly within the limits of agreement, with good agreement and minimal bias between the two methods across most junction geometries. The outcomes obtained for geometry (ii) support our hypothesis that the HMM molecules attached to the channel walls effectively narrow down the channel width, and the filaments can penetrate the outgoing channels once they reach the junction. The results for geometry (iii) act as a control to our hypothesis where the fit to the experimental results was generally poor (as shown in Supplementary section 3 and Figure S3).

However, even for version (ii), with HMM layers included on the entrance channel) there were some notable outliers which lie outside the limits of agreement (labelled in Figure 6). Particularly, the $F_0$ junction design exhibited significantly higher error rates in the simulations than in experiments. Analysis of the generated model movie (see Supplementary video S3) suggests that this discrepancy stemmed from many filaments taking sharp turns and colliding with the offset junction corner. We hypothesize that this behavior arises due to the presence of sharp corners in the simulated geometries, while in actual experiments, rounded corners are expected due to imperfect reproduction of the original design (inset Figure 6(A)) by the electron-beam lithography. To investigate this possibility further, we incorporated curved corners in the simulations. (see details in Supplementary section 5). However, these adjustments failed to effectively reduce the observed errors shown in Supplementary Figure S4 suggesting remaining limitations of the current simulations possibly because we only simulate the filament tips, ignoring tension in a filament that is permanently curved because it moved through a kink in a channel. This finding suggests that factors like rounded corners and other subtle elements, which are not entirely captured in our simulations, have significant influence in faithfully replicating the experimental conditions.

Another notable outlier in geometry (ii) is observed in the $F_{25}$ design, where the experimental error rates were smaller than those predicted by the simulation with geometry (ii). Detailed examination of counting the errors in the experiments for this design reveals that the errors predominantly occur when filaments traverse from entrance b to exit d within all the pass junctions (see Figure 1). The junction design is symmetric, so similar errors are expected from entrance a to exit c as from entrance b to exit d. This asymmetric error is a clear indication of some effect of fabrication in the experiments. Intriguingly, these specific errors were absent in the simulated movie (see Supplementary video S4), indicating the existence of critical factors within the experimental setup that are not accounted for in the simulation. The findings indicate that the optimal thickness for the HMM layer falls within the range of 25-30 nm, as this range exhibits the highest level of agreement. Furthermore, it was shown that the edge stickiness factor appears to exert a relatively minor influence on the simulations (see Table S1 in supplementary information).

**Conclusion**

In summary, our optimization process allowed us to identify a pass junction geometry with $E <$ 1% for the actin-myosin system an improvement by more than a factor of two compared to previous work[3]. Comparison of simulations and experimental data indicate that the key factors associated with low error rate were (i) low channel width (ii) small offset (iii) large funnel width. However, our analysis also indicates that further drastic reduction in errors is unlikely to be feasible when using two-dimensional pass junctions. As a potential approach to completely eliminate errors , the implementation of 3D junctions that restrict the movement of agents to designated paths would be useful[20].

A key finding is that the incorporation of the heavy meromyosin (HMM) layer into the simulations significantly improved agreement with experiment. However, discrepancies between simulation and experiment remain and we tentatively attribute this to the presence of rounded corners in the experimental junctions that are not captured by present simulations. Moreover, other factors, including filament length and interference between filaments, have been observed to influence filament behavior but are not adequately simulated by the current model. Nevertheless, simulations provide a comprehensive understanding of filament behavior which will help design the pass junctions.

**Methods**

**Fabrication of NBC devices.** NBC devices were created using Si (100) substrates measuring 10 × 8 mm². A 75 nm thick $SiO_2$ layer was formed on the Si substrate through thermal oxidation. The substrates underwent oxygen plasma ashing at 5 mbar for 60 seconds and were subsequently cleaned in acetone and isopropanol for 3 minutes each using an ultrasonic bath at room temperature. The cleaned substrates were then dried under a flow of nitrogen. A spin-coating process was employed to apply a layer of approximately 400 nm thickness of CSAR 62 (Allresist GmbH, Strausberg, Germany), which was subsequently baked on a hotplate at 180°C for 120 seconds. The network pattern was defined using electron beam lithography (Raith Voyager) with a dose of 250 μC/cm². The CSAR 62 chips were immersed in amyl acetate developer for 90 seconds and rinsed in IPA for 30 seconds, followed by drying under nitrogen flow. Following the development, the chips were subjected to oxygen plasma ashing at 5 mbar for 45 seconds. This process rendered the CSAR 62 hydrophilic, creating selective guiding between the exposed channel floors and the resist walls, thereby preventing actin attachment. To achieve the desired surface chemistry for supporting actin motility on exposed $SiO_2$ surfaces, the samples were silanized with trimethylchlorosilane in a controlled chamber at 200 mbar.

**Chemicals.** All chemicals used in this study were of analytical grade and purchased from Sigma Aldrich (now Merck) except for Rhodamine Phalloidin which was purchased from Thermo Fisher Scientific (Cat. No.: R415). Parafilm-M was purchased from Sigma Aldrich (now Merck, Cat. No.: P7793). Microscope immersion oil, type-F was purchased from Nikon Instruments (Cat. No.: MXA22192), silicone vacuum grease from Beckman (Prod. No.: 335148), 1 ml syringes and 0.4 × 20 mm needles from B. Braun (Omnifix-F Solo and Sterican).

**Ethical statement.** To obtain actin and myosin motor proteins, rabbits were sacrificed following the standard procedure and guidelines provided by the Regional Ethical Committee for Animal Experiments in Linköping, Sweden (reference numbers: 73-14 and 17088-2020).

**Protein purification.** Myosin motor protein was purified from rabbit fast skeletal muscle[21] and heavy meromyosin (HMM) was prepared by limited proteolysis using chymotrypsin[22]. Prepared HMM was flash-frozen in liquid nitrogen at a concentration ≥ 500 µg/ml and stored at -80 °C until use. Actin was isolated and purified from rabbit back muscle as described previously[23]. Actin monomers were allowed to polymerize by increasing the salt concentration and further incubation at 4 °C. Prepared filaments were aliquoted, flash-frozen in liquid nitrogen, and stored at -80 °C. The concentration and purity of the proteins were evaluated using absorbance spectrophotometry and sodium dodecyl sulfate-polyacrylamide gel electrophoresis (SDS-PAGE). To label actin filaments with rhodamine-phalloidin, an aliquot of actin filaments was allowed to slowly defrost on ice when filaments were diluted to reach a final concentration of 4 µM in actin labeling buffer containing 6 µM of rhodamine phalloidin dye ([actin]:[dye] = 1:1.5).

**Buffer and solutions used in IVMA.** All buffers and solutions were prepared in a degassed low-ionic strength solution (LISS). For degassing the LISS was transferred to a conical flask with a side outlet connected to the air suction line. The flask was covered from the top using a rubber stopper to create low air pressure ($\approx$ -0.8 bar) during suction. For an approximate LISS volume of 20 ml, degassing was done for at least 40 min. LISS was composed of 10 mM 3-(N-morpholino) propane sulfonic acid (MOPS), 1 mM magnesium chloride ($MgCl_2$), and 0.1 mM potassium ethylene glycol-bis(β-aminoethyl ether)-N, N, N', N'-tetraacetic acid ($K_2EGTA$), with final ionic strength of 15 mM and pH of 7.4. Wash buffer was prepared using LISS with the addition of 50 mM KCl and 1 mM DTT. Assay buffer of ionic strength 60 mM was prepared using LISS with addition to the LISS solution of 10 mM DTT, 45 mM KCl, 2.5 mM creatine phosphate (CP), 0.2 mg/ml creatine phosphokinase (CPK), 1 mM magnesium adenosine triphosphate (MgATP) and, finally, an oxygen scavenger mixture (GOC): 3 mg/ml glucose, 0.1 mg/ml glucose oxidase, 0.02 mg/ml catalase. The prepared assay buffer was transferred further to a 1 ml syringe fitted with a hypodermic needle. This syringe was then covered with aluminum foil and kept on ice.

**Flow cell assembly.** To construct flow cells, the chip with the device(s) was affixed face down to untreated microscope coverslips (thickness # 0, 24 × 60 $mm^2$, Menzel-Gläser, Braunschweig, Germany) by securing it with two strips of parafilm stretched to the tension limit. The gap between the two spacers was approximately 8–10 mm. The flow cell was briefly heated on the hot plate to melt the Parafilm and further secure the chip on the coverslip.

*In vitro* **motility assay.** The IVMA was performed[6,24] in flow cells assembled as described above. The flow cells were incubated in a step-wise manner: starting with heavy meromyosin (HMM in wash buffer; 120 $\mu$g $ml^{-1}$) for 5 min, followed by bovine serum albumin (BSA in wash buffer; 1 mg $ml^{-1}$) for 2 min, 1 volume rinsing with wash buffer, then incubation with blocking actin (non-fluorescent actin filaments in wash buffer; 0.5-1 $\mu$M) for 2 min, followed by magnesium adenosine triphosphate (MgATP in wash buffer; 1 mM) for 2 min, 2 volumes rinsing with wash buffer, further incubation with rhodamine-phalloidin labeled actin filaments (in wash buffer; 15 nM), followed by 2 volume rinsing with wash buffer and finally, an assay buffer was added. Sealing of the flow cell was performed after all IVMA incubation steps. For this purpose, we used silicon high vacuum grease that was applied on both open ends of the flow cell.

**Imaging.** An inverted fluorescence microscope (Axio Observer.D1, Zeiss) was used for the observation equipped with a 63× oil immersion objective. For the excitation of rhodamine fluorescence, a short-arc mercury lamp (103 W/2, from OSRAM) was used together with an appropriate filter cube (Cy3). Image acquisition was performed using an electron multiplying charge-coupled device (EMCCD) camera (C9100-12PHX1, Hamamatsu Photonics). Image sequences (videos) were recorded with a time exposure of 200 ms yielding a frame rate of 4.95 frames/second. The pixel size was 0.24 × 0.24 $\mu m^2$ and the image depth was 16-bit. All observations were performed at 25 ±1 °C with the help of a ring-shaped objective heater (Temp controller 2000-2, Pecon). Analysis of the recorded images (i.e. counting correct and incorrect filaments at the pass junction) was done manually using ImageJ[25].

**AUTHOR INFORMATION**

**Corresponding Author**

*Heiner Linke- NanoLund, Box 118, Lund University, 22100 Lund, Sweden and Solid-State Physics, Box 118, Lund University, 22100 Lund, Sweden. Email: heiner.linke@ftf.lth.se

## Acknowledgements

This work was funded by European Union Seventh Framework FET Programme under contract 732482 (Bio4Comp); The Swedish Research Council (project number 2020-04226) and NanoLund.

# Supplementary material

# Error-rate reduction in network-based biocomputation

*Pradheebha Surendiran[1,2], Marko Ušaj[3], Till Korten[4], Alf Månsson[1,3], Heiner Linke[1,2*]*

1 NanoLund, Box 118, Lund University, 22100 Lund, Sweden

2 Solid State Physics, Box 118, Lund University, 22100 Lund, Sweden

3 Department of Chemistry and Biomedical Sciences, Linnaeus University, Kalmar, Sweden

4 B CUBE - Center for Molecular Bioengineering, Technische Universität Dresden, D-01307 Dresden, Germany

**Supplementary videos**

***Supplementary video 1:*** The movie created using the simulation of the filament movement in the guiding structure of $C_{200}$

***Supplementary video 2:*** The movie created using the simulation of the filament movement in the guiding structure of $F_{110}$

***Supplementary video 3:*** The movie created using the simulation of the filament movement in the guiding structure of $F_0$

***Supplementary video 4:*** The movie created using the simulation of the filament movement in the guiding structure of $F_{25}$

**Supplementary information**

*1. Simulation model*

The filament tip is simulated because the tip finds the next motor and moves along that path. The path of the filament is generated using normally distributed angular updates in sliding direction $\Delta\theta$ depending on the persistence length $(L_p)$ and distance between time steps.

$$\text{Angular change per step: } \Delta\theta = \sqrt{\frac{v_f * \Delta t}{L_p}} \qquad (1)$$

Here, $v_f$ is the sliding velocity, $\Delta t$ is the time interval (10 ms) between updates in sliding direction, and $L_p$ is the actin filament persistence length[1,2]

To simulate the sliding motion of filaments, changes in direction are randomly generated from a normal distribution using MATLAB functions. If the filaments move beyond the defined boundaries of the motility-supporting zone because of these changes, they are directed along the boundary. In this study, we adopt the method shown in Figure 1(A) where we first simulate filament movement from its initial position in a motility-supporting zone, assuming that the wall is not present, with a standard deviation for angular change given by (1). If the resulting position (b) falls outside the boundary of the motility supporting region, the filament is redirected to the wall at a new position (c) that is consistent with a sliding distance along the wall of $v_f \Delta t$ from the starting position (a). The edgestickyness parameter is the the probability of the filament to get stuck on the wall and turn around. The probability increases with the absolute sine of the angle at which the filament hits the wall. Stuck filaments turn around. Sticking probability depends on collision angle: $p = sin(\theta)^4 \cdot s$

Where p: sticking probability; $\theta$: collision angle; s: sticking parameter

The simulation code: (a) creates a guiding structure from the design file of the networks (b) simulates the filaments by giving properties of them such as speed, persistence length, interval (c) visualizes the simulated results by creating movies and heatmaps of the filament paths.

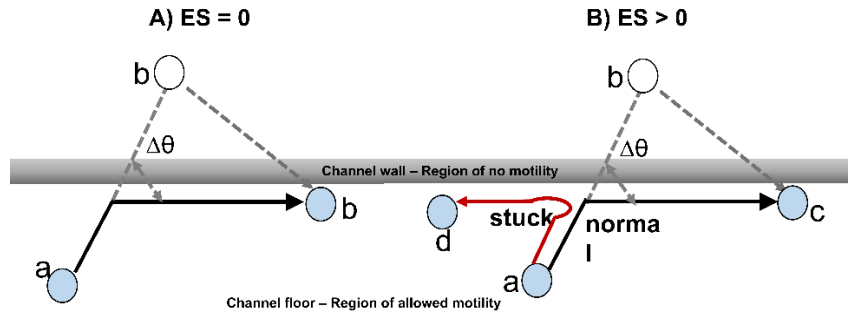

**Figure S1.** Working of the simulation model

.

## 2. Simulation of modified design by reducing one entrance channel

When the filaments move from the entrance channel to the junction, the filaments can penetrate the motor layer in all the three outgoing channels as they will enter with the leading tip. This behavior is true for both the entrance channels. In simulation, this means that when one of the channels is considered as incoming channel where filaments cannot penetrate, that channel is reduced in width whereas the other three channels remain in the original width, since these channels can be penetrated. This means that the designs where both entrance channels are reduced are flawed and hence, we need to reduce one channel at a time like the half design.

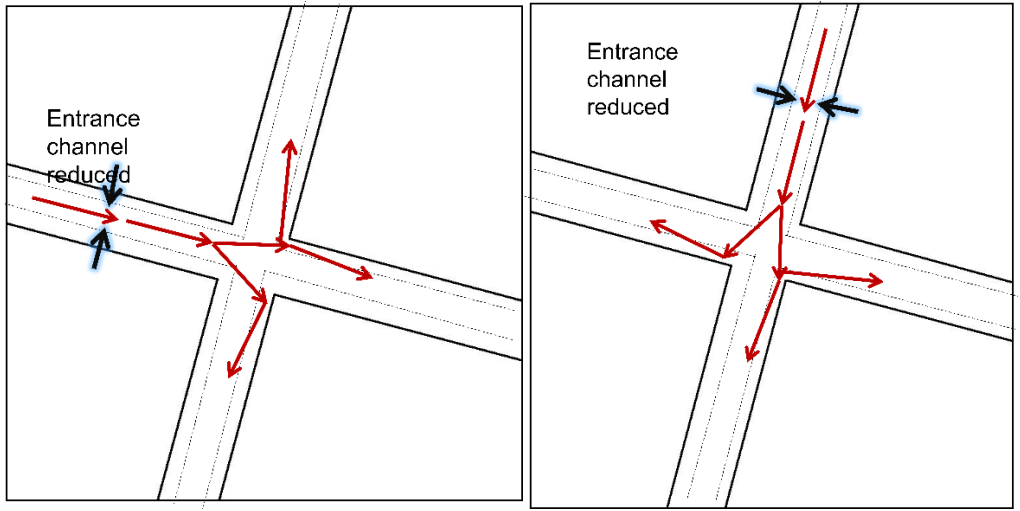

**Figure S2.** Movement of the filaments is represented by red arrows. The leading tip enters the junction and penetrates the outgoing channels. The black arrows show the entrance channels reduced to the width of the HMM layer, represented by the dotted lines.

## *3. Modified design with HMM layer in all channels*

To support the hypothesis outlined in section 2, we conducted additional simulations involving designs in which all the channels were altered (design (iii)), considering the presence of an HMM layer. In these simulations, we assumed that the filament would be unable to penetrate the HMM layer as it entered the junction, serving as a control. Figure S3 illustrates limited alignment between the error rates obtained from the experimental data and the simulation results.

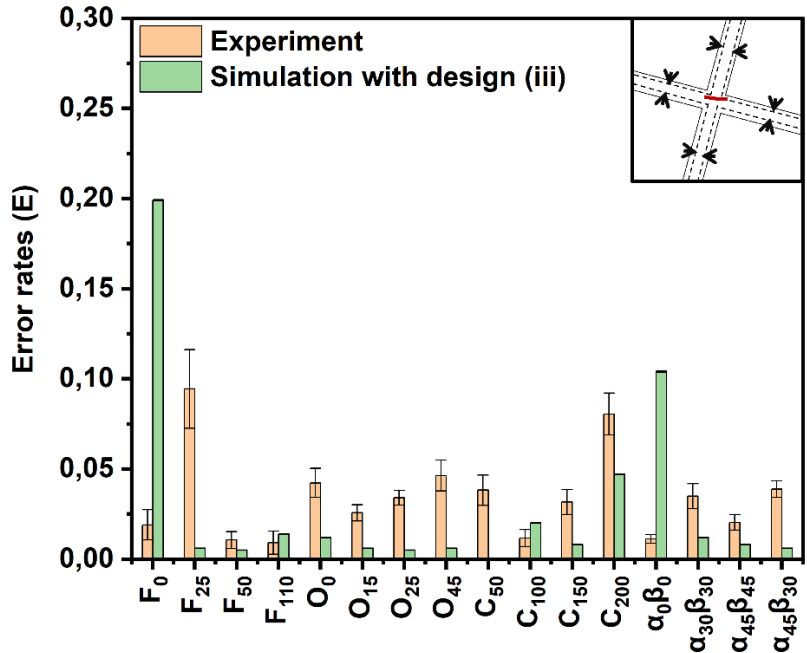

**Figure S3.** Comparison of experimental and simulation error rates. Simulation results of design (iii) with all channel widths reduced as shown by black arrows in inset to account for a heavy meromyosin HMM layer on the channel walls (dashed lines in inset) and filament is shown in red color.

*4. Statistical evaluation*

To explore the optimal combination of parameters, simulations were conducted with varying HMM thicknesses (ranging from 25 to 35 nm) and Edgestickiness values (ranging from 0 to 0.16). To provide quantitative assessment of the agreement between the simulation and experimental data, Bland-Altman plot was employed. A Bland-Altman plot is a graphical method used in statistics to assess the agreement between two different measurements. It is particularly useful for comparing the agreement between two methods. The Bland-Altman plot consists of

**Y-Axis:** On the y-axis, plot the differences between the measurements obtained from the two methods. It is measured as Experimental error rates – Simulation error rates.

**X-Axis:** On the x-axis, plot the average or mean of the measurements obtained from the two methods. This is usually represented as (Experimental error rates + Simulation error rates) / 2.

**Horizontal Lines:** The upper line represents the mean difference plus 1.96 times the standard deviation of the differences, while the lower line represents the mean difference minus 1.96 times the standard deviation of the differences. These lines indicate the limits of agreement (LoA) or the range within which most of the differences between the methods are expected to fall.

**Scatter Plot:** The individual data points are then plotted as dots on the graph, with their x-coordinate being the average of the two measurements and the y-coordinate being the difference between the two measurements.

**Table S1.** Mean and limits of agreement of the Bland-Altman plot for range of HMM thickness and edgestickiness

| Edgestickiness | HMM_25 nm | | | HMM_30 nm | | | HMM_35 nm | | |
| --- | --- | --- | --- | --- | --- | --- | --- | --- | --- |
| | Mean | Upper LoA | Lower LoA | Mean | Upper LoA | Lower LoA | Mean | Upper LoA | Lower LoA |
| **0** | 0,0065 | 0,0823 | -0,0693 | 0,0098 | 0,0840 | -0,0645 | 0,0156 | 0,0963 | -0,0652 |
| **0.04** | 0,0030 | 0,0725 | -0,0665 | 0,0052 | 0,0826 | -0,0723 | 0,0102 | 0,0904 | -0,0701 |
| **0.08** | 0,0008 | 0,0703 | -0,0688 | 0,0020 | 0,0774 | -0,0734 | 0,0076 | 0,0878 | -0,0726 |
| **0.12** | -0,0029 | 0,0672 | -0,0730 | **-0,0006** | 0,0762 | -0,0773 | 0,0050 | 0,0839 | -0,0740 |
| **0.16** | -0,0055 | 0,0638 | -0,0749 | -0,0023 | 0,0709 | -0,0754 | 0,0020 | 0,0809 | -0,0769 |

*5. Testing rounded corners in simulation*

The design $F_0$(straight junction with an offset) was observed as an outlier. The errors in this design were observed to be mainly due to filaments hitting the junction corner and taking turns in the simulation. Small corner modifications to the design were made to translate the rounded corners in the experiment to simulations. Based on this, simulations were done with different ES and HMM thicknesses to see if that influences the resulting error rates, still the error rates were in the range of 12-15% like the current simulation data. Pixel values affecting the resolution of the simulations were also varied to see if that affects the translation of small changes in designs but there was no improvement in the error rates.

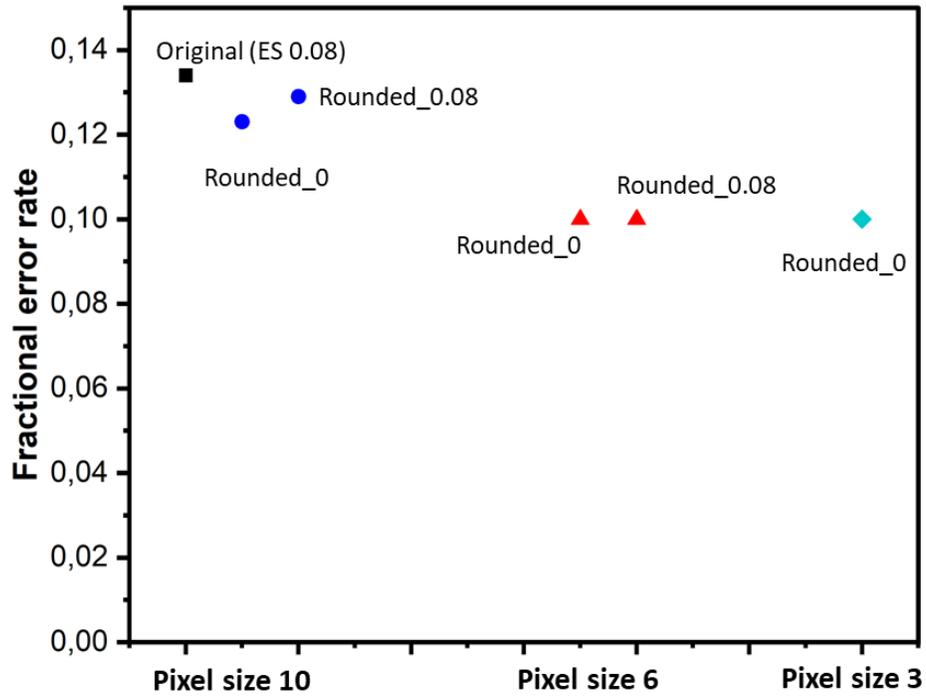

**Figure S4.** Error rates of $F_0$ design. The black shows the error rate from the current simulation. Indigo, red and blue represent the error rates of modified design with rounded corners for different pixel values of 10, 6 and 3 respectively. 0 and 0.08 are the ES values used for the respective values.